\documentclass[prl,twocolumn]{revtex4}
\usepackage{graphicx}
\usepackage{amsmath}
\begin{document}

\title{Full Counting Statistics of a Superconducting Beam Splitter}

\author{J. B\"orlin}
\author{W. Belzig}
\author{C. Bruder}
\affiliation{Departement f\"ur Physik und Astronomie, Klingelbergstr. 82, 4056
 Basel, Switzerland}
\date{30 April 2002; published in Phys. Rev. Lett. {\bf 88}, 197001 (2002)}
\begin{abstract}
  We study the statistics of charge transport in a mesoscopic
  three-terminal device with one superconducting terminal and two
  normal-metal terminals. We calculate the full distribution of
  transmitted charges into the two symmetrically biased normal
  terminals.  In a wide parameter range, we find large positive
  crosscorrelations between the currents in the two normal arms. We
  also determine the third cumulant that provides additional
  information on the statistics not contained in the current noise.
\end{abstract}



\maketitle

The number of charges transfered in a transport process fluctuates due to
quantum-mechanical uncertainty and statistics. Therefore, the outcome of a
current measurement accumulated over some time period $t_0$ is in general
described by a probability $P(N)$, where $N$ is the total number of charges
transfered. $P(N)$ is called the \textit{full counting statistics} (FCS) of
the transport process \cite{levitov:96}. The first two moments of the FCS are
related to the average current and the current noise and are accessible to
present experimental techniques. Higher-order correlations are likely to be
measured in the future. Several schemes to measure either higher correlators
or the full distribution have been proposed recently
\cite{levitov:96,belzig:01-2,levitov:01,kindermann:01,pumps,beenakker}.

The current noise, i.~e., the second moment of the FCS, is of
particular interest. It can be used as a diagnostic tool to probe the
nature and the quantum statistics of the charge carriers
\cite{blanter} and the existence of entanglement \cite{loss}. For
superconductor(S)-normal metal(N) heterostructures, a doubling of the shot
noise in comparison to the normal case was predicted \cite{2e} and
measured in diffusive heterostructures \cite{2e-measure}. Recent
calculations taking into account the proximity effect in such
structures \cite{belzig:01-1} are in good agreement with experimental
results \cite{kozhevnikov}. Multi-terminal SN structures have been
suggested to produce entangled electron pairs \cite{recher,lesmarblat}.

So far, crosscorrelations, i.~e., current correlations involving different
terminals, were measured only in normal single-channel heterostructures
\cite{hantwissexp}. These have confirmed the prediction \cite{buettiker:91}
that current crosscorrelations in a fermionic system are always negative. To
our knowledge, there is no measurement of crosscorrelations in a system with
superconducting contacts up to now. Theoretically, positive crosscorrelation
with a single-channel beam splitter for Andreev pairs injected from a
superconductor have been predicted \cite{martin}. In a setup in which
crosscorrelations between a normal lead and a tunneling probe are considered,
the sign of the correlations was found to depend crucially on the sample
geometry \cite{gramesbacher}. A numerical study found positive
crosscorrelations in a three-terminal device with a few channels with
ferromagnetic contacts \cite{taddei2}.

In this Letter we find the full counting statistics of a many-channel
beam splitter that divides a supercurrent in two normal quasiparticle
currents. We calculate the distribution of the transmitted charges taking
the proximity effect into account. For comparison we also calculate
the FCS for the case in which the superconducting terminal is replaced by a
normal one.

\begin{figure}[b]
 \begin{center}
 \includegraphics[width=4.5cm,clip=true]{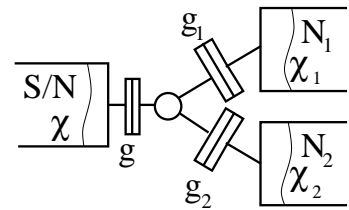}
 \caption{Three-terminal beam-splitter.
   One superconducting or normal terminal (S or N) and two normal
   terminals ($N_1$ and $N_2$) are tunnel-coupled by conductances $g$,
   $g_1$ and $g_2$ to a common central node .  A current is passed from
   S/N into the two normal leads that are kept at the same voltage.
   Ideal passive charge counters are indicated by the counting fields
   $\chi$, $\chi_1$ and $\chi_2$.}
 \label{fig:system}
 \end{center}
\end{figure}

The setup of our three-terminal device with one superconducting and
two normal-metal terminals is shown in Fig.~\ref{fig:system}. All
three terminals are connected by tunnel junctions to a small
normal-metal island. We assume the island to be large enough, that we
can neglect charging effects, and small enough that we can treat the
Green's functions of the island as constant. Thus, we are restricted
to energies below the Thouless energy of the island. The system is
then appropriately described by the circuit theory of mesoscopic
transport \cite{yuli:9499-1}, within which the counting
statistics is easily accessible
\cite{yuli:99-2,belzig:01-1,belzig:01-2}.

The circuit elements that are important for our case are normal, resp.
superconducting terminals and tunnel connectors. The terminals are described
by quasiclassical Green's functions, which depend on the type of the terminal
(N or S), chemical potential, temperature $T$ and a counting field
$\chi$. We assume zero temperature and a symmetric bias at the two normal
terminals. Since we are interested in correlations between currents in
different terminals, we introduce different counting fields. The
voltage is chosen such that $eV\ll\Delta$. Charge transport (at $T=0$)
occurs then only in the interval $|E|\leq eV$ and we need to consider only
this energy interval below.

The Green's functions of the two normal terminals are then given by
\begin{equation}
 \label{eq:normal-greensfunc}
 \check G_{1,2} = e^{i\chi_{1,2}\check \tau_K/2} \check G_{N}
 e^{-i\chi_{1,2}\check \tau_K/2}\,,
\end{equation}
where $\check G_N$ is the same for both normal terminals. At zero
temperature $\check G_N=\hat \sigma_3 \bar \tau_3 + (\bar \tau_1+i\bar
\tau_2)\hat 1$ for $|E|\leq eV$ and $\check G_N=\hat \sigma_3
\bar \tau_3 + \mbox{sgn}(E)\hat \sigma_3(\bar \tau_1+i\bar \tau_2)$ for
$|E|>eV$.  Here $\hat \sigma_i (\bar \tau_i)$ denote Pauli matrices in
Nambu(Keldysh)-space. The counting rotation matrix is $\check
\tau_K=\hat \sigma_3\bar \tau_1$. The superconducting terminal in
equilibrium is characterized by $\check G_S=\hat\sigma_1\bar 1$
and a counting field $\chi$, 
that enters as in (\ref{eq:normal-greensfunc}).

If one node is connected to $M$ terminals by means of tunnel connectors, one
can find a general form of the FCS, \textit{i.e.} the probability
$P(N_1,\ldots,N_M)$ that $N_{1(2,\ldots,M)}$ charges are counted in terminal
$1(2,\ldots,M)$.  The unknown Green's function of the central node is denoted
by $\check G_c$. The matrix currents into the central node are given by
$
 \check I_k = \frac{g_k}{2} \left[ \check G_c,\check G_k \right]
$,
where the index $k=1,\ldots,M$ labels the terminals and $g_k$ is the
conductance of the respective junction. The Green's function of the central
node is determined by matrix current conservation on the central node, reading
$ \sum_{k=1}^M \check I_k = \frac12 \left[ \sum_{k=1}^M g_k \check G_k ,
  \check G_c \right]=0$.  Employing the normalization condition $\check
G_c^2=1$, the solution is
\begin{equation}
 \check G_c = \frac{\sum_{k=1}^M g_k \check G_k}{
 \sqrt{\sum_{k,m=1}^M g_kg_m
 \left\{\check G_k,\check G_m\right\}/2}}\,.
\end{equation}
To find the cumulant-generating function (CGF) $S$ of $P(N_1,\ldots,N_M)$ we
integrate the equations $(-it_0/e) \partial S(\chi_1,\ldots,\chi_M)/\partial
\chi_k = \int dE \mbox{Tr}\check\tau_K\check
I_{k}/8e$ \cite{yuli:01-1}.  We obtain
\begin{equation}
 \label{eq:cgf-general}
 S(\chi_1,\ldots,\chi_M)
 =- \frac{t_0}{e}\int \frac{dE}{2} \mbox{Tr} 
 \sqrt{\sum_{k,m=1}^M \frac{g_kg_m}{2}
 \left\{\check G_k,\check G_m\right\}}\,.
\end{equation}
This is the general result for an M-terminal geometry in which 
all terminals are tunnel-coupled to a common node. 

We now evaluate Eq.~(\ref{eq:cgf-general}) for our three terminal
setup.  Introducing $p_i=2 g g_i /(g^2+(g_1+g_2)^2)$ we find
\begin{widetext}
\begin{equation}
 \label{eq:cgf}
 S(\chi_1,\chi_2,\chi) = -\frac{Vt_0\sqrt{g^2+(g_1+g_2)^2} }{\sqrt 2 e}
 \sqrt{1+ \sqrt{1+
 \left(p_1 e^{i(\chi_1-\chi)}+p_2e^{i(\chi_2-\chi)}\right)^2-(p_1+p_2)^2}}\,.
\end{equation}
\end{widetext}
This result for the cumulant-generating function incorporates all
statistical transport properties for our present setup. The inner
argument contains counting factors for the different possible
processes. A term $\exp(i(\chi_k+\chi_l-2\chi)-1)$ corresponds to an
event in which two charges leave the superconducting terminal and one
charge is counted in terminal $k$ and one charge in terminal $l$. The
prefactors are related to the corresponding probabilities. For
instance, $p_1$ is proportional to the probability of a coherent
tunneling event of an electron from the superconductor into terminal 1. A
coherent pair-tunneling process is therefore weighted with
$p_1^2$. This is accompanied by counting factors which describe
either the tunneling of two electrons into terminal 1(2) 
[counting factor $\exp(i2(\chi_{1(2)}-\chi))-1$] 
or tunneling into different terminals 
[counting factor $\exp(i(\chi_1+\chi_2-2\chi))-1$].
The double square-root function shows that these different processes
are non-separable.

It is interesting to compare Eq.~(\ref{eq:cgf}) with the case in which
the superconductor is replaced by a normal metal.  The resulting CGF is
\begin{equation}
 \begin{array}[t]{l}
 S^N(\chi_1,\chi_2,\chi)=-\frac{Vt_0}{2e}(g+g_1+g_2)\times\\\\
 \quad\sqrt{1+ p^N_1 \left(e^{i(\chi_1-\chi)}-1\right) 
  + p^N_2 \left(e^{i(\chi_2-\chi)}-1\right)}\,,
 \end{array} \label{eq:cgf-normal}
\end{equation}
where $p^N_{1(2)}=4 g g_{1(2)}/(g+g_1+g_2)^2$. Thus, one of the
square roots in Eq.~(\ref{eq:cgf}) can be attributed to the multiple
tunnel-junction geometry, which is already present in the normal
configuration. The second square root in the CGF
for the superconducting case must then be due to the proximity
effect.

We now evaluate some average transport properties of the S$\mid$NN-system and
compare them to the N$\mid$NN-case. The currents into the different terminals
are obtained from derivatives of the CGF: $I_k=(-ie/t_0) \partial
S/\partial\chi_k|_{\chi_1=\chi_2=\chi=0}$. The trans-conductances
$G_k=I_k/V$ into terminal $k$ $(=1,2)$ are then given by
\begin{eqnarray}
 G_k^S = \frac{g^2g_k(g_1+g_2)}{
 \left(g^2+\left(g_1+g_2\right)^2\right)^{3/2}}&,&
 G_k^N = \frac{gg_k}{g+g_1+g_2}\,.
\end{eqnarray}
The superscript $S(N)$ denotes the S$\mid$NN(N$\mid$NN)-case.  Noise and
crosscorrelations are obtained from second derivatives of the CGF, i.~e.,
$P_{kl}^I=(2e^2/t_0) \partial^2 S(\chi_1,\chi_2,\chi)/\partial \chi_k\partial
\chi_l|_{\chi_1=\chi_2=\chi=0}$. We define Fano factors $F_{kl}=P^I_{kl}/2eI$,
and we denote the Fano factor of the total current with
$F=F_{11}+F_{22}+2F_{12}$. We also calculate the third cumulant of the total
charge transfer (normalized to the Poisson value) $C_3=(ie/It_0)\partial^3
S(0,0,\chi)/\partial\chi^3|_{\chi=0}$.  The results in the superconducting
case are
\begin{eqnarray}
 \label{eq:fss}
 F_{12}^S = \frac{g_1g_2}{(g_1+g_2)^2}(1-5x^2)&,&F^S = 2- 5x^2\,,
 \\\nonumber
 C^S_3= 4-30 x^2+63 x^4 &,& x=\frac{g (g_1+g_2)}{g^2 +(g_1+g_2)^2}\,.
\end{eqnarray}
In the N$\mid$NN case, on the other hand, we find
\begin{eqnarray}
 \label{eq:fnn}
 F_{12}^N = -\frac{g_1g_2}{(g_1+g_2)^2}x_N&,& F^N = 1-2x_N\,, \\\nonumber
 C_3^N=1-6x_N+3x_N^2 &,& x_N=\frac{g(g_1+g_2)}{(g+g_1+g_2)^2}\,.
\end{eqnarray}
\begin{figure}[tbp]
 \begin{center}
 \includegraphics[width=7cm,clip=true]{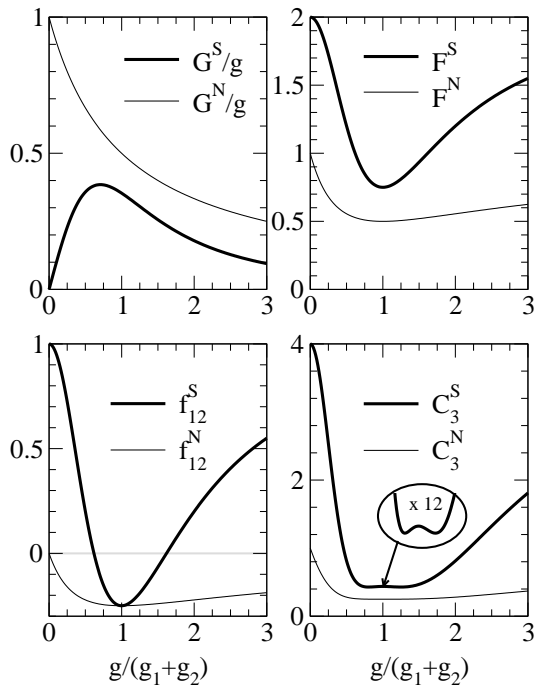}
 \caption{Conductance, Fano factors, crosscorrelations and third 
   cumulant of the beam splitter. The thick lines correspond to the
   S$\mid$NN-case and the thin lines to the N$\mid$NN-case. The conductance
   (upper-left panel) in the superconducting case shows a maximum around
   $g=g_1+g_2$. In the normal state, the conductance varies between $g$
   and $g_1+g_2$. In both cases, the current noise (upper-right panel) shows a
   suppression around $g=g_1+g_2$ as compared to the limiting values of 2 in
   the S$\mid$NN-case and 1 in the N$\mid$NN-case.  Large positive
   crosscorrelations occur in the superconducting case (lower-left panel),
   whereas they are always negative in the normal case. Around $g=g_1+g_2$,
   the superconducting crosscorrelations become negative. Note, that what is
   plotted here is $f_{12}^{S/N}= F_{12}^{S/N}(g_1+g_2)^2/g_1g_2$. The third
   cumulants (lower-right panel) are always positive. Around $g=g_1+g_2$ they
   are strongly suppressed. In the S$\mid$NN-case, $C_3^S$ has a 
   double-minimum here, as shown in the blow-up.}
 \label{fig:noise1}
 \end{center}
\end{figure}
All other Fano factors can be deduced from $F_{12}$ and $F$ using the
relations $\sum_k F_{kl}=0$ and $F_{kl}=F_{lk}$. The transport properties are
summarized in Fig.~\ref{fig:noise1}. In the figure the crosscorrelations are
plotted as $f_{12}=F_{12} (g_1+g_2)^2/g_1g_2$. Most remarkably, the
crosscorrelations $F_{12}^S$ are positive if $x$ is small, whereas $F_{12}^N$
is always negative in the normal state. Here, the Fano factor $F^S$($F^N$) is
close to $2$($1$). Going to the regime $g \approx (g_1+g_2)$ suppresses the
Fano factor $F^{S(N)}$ below 2(1) and leads to negative crosscorrelations
$F_{12}^S$. In the limiting case $g = (g_1+g_2)$ the Fano factors are
$F^{S}=3/4$ and $F^N=1/2$, and the crosscorrelations are $F_{12}^S=F_{12}^N =
-g_1g_2/4(g_1+g_2)^2$. The third cumulant is always positive, but shows a
strong suppression around the resonant conductance ratio $g = (g_1+g_2)$. In
the limit of small $x$ ($x_N$) the third cumulant is $4$ ($1$), corresponding
to the effective charge squared transfered in a tunneling process
\cite{levitov:01}. However, the variation with $g/(g_1+g_2)$ in the
S$\mid$NN-case is more pronounced than in the N$\mid$NN-case.

As an interesting side remark we point out that $F_{12}^S=0$ and $F^S=1$ for
$x^2=1/5$. This looks like a signature of uncorrelated charge transfer in
units of $e$. However, the third cumulant $C_3^S=13/25$ differs from the
corresponding value for uncorrelated $1e$-charge transfer, viz., $C_3=1$.
Thus, higher correlations show that the charge transfer is still correlated.

We briefly discuss the influence of an asymmetry $g_1\neq g_2$ of the beam
splitter. The crosscorrelations are reduced, both in the S$\mid$NN and in the
N$\mid$NN case. However, the positive crosscorrelations in the superconducting
state persist for all values of the asymmetry. Cumulants of the total charge
transfer like the conductance, $F^{S,N}$ and $C_3^{S,N}$ are independent of
this asymmetry.

Using the CGF from Eq.~(\ref{eq:cgf}), we can identify the physical
processes leading to our previous results. We have seen from
(\ref{eq:fss}) that positive crosscorrelations are found if
$g/(g_1+g_2)$ is not close to 1. Then, $p_{1,2}\ll 1$ and we can expand
Eq.~(\ref{eq:cgf}) in $p_{1,2}$. Dropping the trivial dependence on
$\chi$, the CGF can be written as
\begin{eqnarray}
 \label{eq:cgf-poisson}
 S(\chi_1,\chi_2) = 
 -\frac{t_0V}{e}\frac{g^2}{(g^2+(g_1+g_2)^2)^{3/2}}\times\qquad\qquad
 \\\nonumber
 \qquad\left( g_1^2 e^{i2\chi_1}+ g_2^2 e^{i2\chi_2}+2 g_1g_2 
e^{i(\chi_1+\chi_2)} \right) \; .
\end{eqnarray}
The CGF is composed of three different terms, corresponding to a
charge transfer of $2e$ either into terminal 1 or terminal 2 (the
first two terms in the bracket) or separate charge transfer into
terminals 1 and 2. According to the general principles of statistics,
sums of CGFs of independent statistical processes are
additive. Therefore, the CGF (\ref{eq:cgf-poisson}) is a sum of CGFs
of independent Poissonian processes. Crosscorrelations are obtained from
derivatives with respect to $\chi_1$ and $\chi_2$. Thus, the first two
terms in (\ref{eq:cgf-poisson}) corresponding to two-particle
tunneling either into terminal 1 or 2 do not contribute. It is only
the last term which yields crosscorrelations, and those are
positive. Poissonian statistics are the statistics of uncorrelated
events, which in our case means all tunneling events are
independent. Thus, a two-particle tunneling event into one of the
normal terminals is not correlated with other tunneling events and
does not contribute to crosscorrelations, but only to the
autocorrelations. The two-particle tunneling into different terminals,
however, is automatically positively crosscorrelated.
The crosscorrelations are therefore positive.

The total probability distribution $P(N_1,N_2)$ corresponding to
(\ref{eq:cgf-poisson}) can be found. It vanishes for odd values of
$(N_1+N_2)$ and for even values it is
\begin{equation}
 \label{eq:distribution-poisson}
 P(N_1,N_2) = 
  \frac{e^{-\frac{\bar N}{2}}
    \left(\frac{\bar N}{2}\right)^{\frac{N_1+N_2}{2}}}{
    \left(\frac{N_1+N_2}{2}\right)!}
  \binom{N_1+N_2}{N_1}
  T_1^{N_1}T_2^{N_2}\,.
\end{equation}
Here we have defined the average number of transfered electrons $\bar N
= t_0 G^S V/e$ and the probabilities $T_{1(2)}=g_{1(2)}/(g_1+g_2)$ that
one electron leaves the island into terminal $1(2)$. If one would not
distinguish electrons in terminals 1 and 2, the charge counting
distribution can be obtained from (\ref{eq:cgf-poisson}) by setting
$\chi_1=\chi_2=\chi$ and performing the integration. This leads to
$P^S_{\textrm{tot}}(N)=\exp(-\bar N/2) (\bar N/2)^{N/2}/(N/2)!$, which
corresponds to a Poisson distribution of an uncorrelated transfer of
electron pairs. The full distribution (\ref{eq:distribution-poisson})
is given by $P^S_{\textrm{tot}}(N_1+N_2)$, multiplied with
a \textit{partitioning factor}, which corresponds to the number of ways how
$N_1+N_2$ identical electrons can be distributed among the terminals 1
and 2, with respective probabilities $T_1$ and $T_2$. Note, that
$T_1+T_2=1$, since the electrons have no other possibility to leave the
island.

In contrast to that, we obtain in the normal case for $t^N_{1,2}\ll 1$
the probability distribution:
\begin{equation}
 \label{eq:normal-poisson}
 P^N(N_1,N_2) = e^{-\bar N_1} \frac{\bar N_1^{N_1}}{N_1!} 
 e^{-\bar N_2}\frac{\bar N_2^{N_2}}{N_2!}\,.
\end{equation}
Here we have abbreviated the average number transfered into terminal $i$
by $\bar N_i$. Thus, the distribution in the normal case is the
product of two Poisson distributions of charge transfers into the two
terminals. In the superconducting case such a factorisation is not
possible.

In conclusion, we have studied the full counting statistics of a
three-terminal device with one superconducting and two normal leads. The
system is biased such that a supercurrent is passed from the
superconductor into the two normal leads, with no net current between
the normal leads. Thus, the device acts as a sort of beam splitter. We
have calculated the full distribution of transmitted charges using the
extended Keldysh-Green's function method fully accounting for the
proximity effect. Our main finding are large positive crosscorrelations
of the currents in the two normal terminals in a wide parameter range.
These should be easily accessible experimentally. These positive
correlations originate from independent Poisson processes of coherent
tunneling of charges into the different terminals. These dominate the
crosscorrelations, since two-particle tunneling into the same lead does
not contribute to the crosscorrelations. We have also calculated the
third cumulant which provides additional information on the current
statistics not contained in the current noise.

\begin{acknowledgments} We would like to thank G. Burkard, D. Loss, and Yu.~V.
  Nazarov for discussions. During the preparation of this manuscript, we
  became aware of similar work by P. Samuelsson and M. B\"uttiker
  \cite{samuelsson}. Our work was supported by the Swiss NSF and the
  NCCR Nanoscience.
\end{acknowledgments}

\end{document}